# Lithography-free electrical transport measurements on 2D materials by direct microprobing

*Patricia Gant[1], Yue Niu[1,2], Simon A. Svatek[1,3], Nicolás Agraït[1,3], Carmen Munuera[4], Mar García-Hernández[4], Riccardo Frisenda[1], David Perez de Lara[1], Andres Castellanos-Gomez[4,]**

[1] *Instituto Madrileño de Estudios Avanzados en Nanociencia (IMDEA Nanociencia), Campus de Cantoblanco, E-28049 Madrid, Spain.*

[2]*National Key Laboratory of Science and Technology on Advanced Composites in Special Environments, Harbin Institute of Technology, Harbin, China*

[3]*Departamento de Física de la Materia Condensada and IFIMAC. Universidad Autónoma de Madrid, Madrid, E-28049, Spain.*

[4]*Instituto de Ciencia de Materiales de Madrid (ICMM-CSIC), Madrid, E-28049, Spain.*

* andres.castellanos@csic.es

ABSTRACT

We present a method to carry out electrical and opto-electronic measurements on 2D materials using carbon fiber microprobes to directly make electrical contacts to the 2D materials without damaging them. The working principle of this microprobing method is illustrated by measuring transport in $MoS_2$ flakes in vertical (transport in the out-of-plane direction) and lateral (transport within the crystal plane) configurations, finding performances comparable to those reported for $MoS_2$ devices fabricated by conventional lithographic process. We also show that this method can be used with other 2D materials.

Isolation of 2D materials by mechanical exfoliation of bulk layered crystals [1] has changed the paradigm of materials science, where typically extremely high quality samples are required to observe intriguing physical phenomena and to understand their origin [2–4]. Until the isolation of graphene [5] and other 2D materials [6–9], those high quality samples were grown by quite sophisticated methods (molecular beam epitaxy, pulsed laser deposition, etc), accessible to only few well-equipped and highly experienced laboratories around the world [2–4,10]. Mechanically exfoliated samples, which can be easily obtained in virtually any materials science laboratory without need of a large investment, have demonstrated a remarkably high quality enabling for the observation of interesting transport phenomena such as Dirac fermions, charge density waves or quantum Hall effect [11–14].

In order to probe the electrical properties of 2D materials or to fabricate proof-of-concept devices, 2D materials have to be typically processed in a nanofabrication facility (i.e. clean room equipped with lithography tools, wet chemistry benches and metal deposition systems). Therefore, although high quality 2D materials can be isolated at almost any laboratory with a minimum infrastructure investment, the access to nanofabrication tools is still an important hurdle for many research groups, especially for those in institutions without background on micro- and nanoelectronics.

The development of a technique to perform electrical measurements on 2D materials that can be easily implemented in laboratories without lithography capabilities can have a strong impact in the community as it would open up a simple route to test as-isolated 2D materials and even to fabricate proof-of-concept devices. That is exactly the main goal of this manuscript.

Previous attempts have been made to make electrical contacts to 2D materials without the use of lithography such as direct metal evaporation through shadow/stencil masks [15–18] and microsoldering [19,20]. Nonetheless the first method requires some infrastructure for metal deposition and the stencil fabrication while microsoldering is non-trivial and requires experience. A desirable alternative is to use a probe station where electrical contacts are directly made on the sample by means of sharp needle-like probes. Conventional probes used in probe stations are made of tungsten and have a diameter >25 µm. These probes are too stiff for direct probing 2D materials and they tend to scratch the flakes making their use unpractical. Nanoscopic versions of probe station systems, able of making gentle contacts to nanomaterials, have been developed but they require the use of highly accurate piezoelectric actuators and they use specialized microscopy systems (AFM, SEM or TEM) to monitor the placement of the probes [21–23].

We have developed a manually actuated microprobe station that allows placing very flexible and floppy electrical probes onto 2D materials without damaging their surface. The electrical probes are made out of polyacrylonitrile (PAN) carbon fibers [24–27] ~1 mm long. We address the reader to the Supporting Information for a characterization of the fiber electrical resistivity and contact resistance [28]. Note that despite the reduced dimension of the fibers (7 µm in diameter, much thinner than a human hair as shown in Supporting Information) they can be easily handled because of their remarkable straightness. The carbon fibers are attached to macroscopic metallic tips by silver loaded adhesive (Figure 1a) and the probes are then mounted on XYZ-axis micrometer stages to accurately position them. We employ a coaxial zoom lens inspection system to place the probes onto the desired part of the flake under study.

An overall view of the experimental setup is shown in Figure 1b. Figure 1c shows one of the two XYZ-stages with a probe arm and the mounted carbon fiber probe. Finally, an image of a $MoS_2$ flake contacted with two carbon fiber probes is shown in Figure 1d, the separation between the two probes (which defines the semiconductor channel length) is in this case ~20 μm. Notice that one can reliably place two carbon fiber probes with a minimum separation of ~10 μm using this setup.

Firstly, we test the performance of the carbon fiber probes by making a single electrical contact to $MoS_2$ flakes deposited on Au substrate, forming a vertical device where the electrical transport happens across the out-of-plane direction (parallel to the c-axis of $MoS_2$). The $MoS_2$ flakes have been produced by mechanical exfoliation with Nitto tape (SPV 224) from a bulk natural crystal (Moly Hill mine, Quebec, QC, Canada) onto a polydimethylsiloxane substrate (PDMS). Then the selected flake is transferred from the PDMS to the Au substrate by an all dry deterministic transfer method [29]. Previously to the transfer, we characterize the flakes on the PDMS by optical transmission and micro-reflectance to determine the number of layers (see Supporting Information and References [30,31] for more details). Alternatively, other optical microscopy based identification methods can be used such as hyperspectral imaging [32,33] or quantitative analysis of the optical microscopy images [34,35]. Figure 2a shows a schematic illustration of the experimental configuration and the electrical connections used to probe out-of-plane transport in $MoS_2$. Figure 2b displays optical images of the $MoS_2$ flake under study deposited on Au and of the flake contacted by the carbon fiber (inset), where the different colors of the flake correspond to different thicknesses. The electrical contact can be made by gently lowering the probe while the current between the sample and the probe is monitored (with a voltage bias applied), an example can be seen in the Supporting Information. Once the contact has been made, a current versus voltage trace (*IV*) can be measured to probe the electrical properties of the flake. Figure 2c shows a set of the *IV*s acquired on different regions of the $MoS_2$ flake with different thicknesses. We observe that the resistance increases exponentially with increasing the number of layers because the MoS$_2$ acts as tunneling barrier [36] whose width depends on the number of layers. Note that by directly contacting the conductive Au substrate (dashed line in Figure 2c) one can determine the fiber resistance ($R$ = 800 ± 50 Ω) and thus subtract it from the *IV*s. Note that the linear Au-carbon fiber probe *IV* also demonstrates that the carbon fibers have metallic conductance without Schottky barrier at the contact.

We now probe another vertical MoS$_2$ device fabricated by transferring the MoS$_2$ flake onto a p-type silicon substrate as depicted in Figure 3a. Previous works showed that a p-n junction is formed when MoS$_2$, a n-type semiconductor, is in contact with p-type Si [37–41]. To fabricate a diode with these materials, we first etched the SiO$_2$ layer from the SiO$_2$/Si substrate and then we transferred a MoS$_2$ flake on the naked Si region, forming a p-n junction. In Figure 3b, we show an optical image of the flake transferred onto the freshly etched Si. The inset shows a transmission mode optical image of the flake before being transferred where the regions of the flake with different number of layers can be distinguished. To perform the measurements, we contacted the monolayer region, which has an area of 900 μm$^2$. Figure 3c shows *IV*s recorded in dark (black curve) and under illumination with a white LED source (blue curve). The dark *IV* displays the distinctive rectifying behaviour of a diode typical of a p-n junction, clearly different from the previous device (MoS$_2$/Au). Upon illumination, the *IV* shows an open-circuit voltage of ~0.15 V and a short-circuit current of ~5 nA, signature of the photovoltaic photocurrent generation mechanism [37–41].

We now turn our attention to the use of two carbon fiber probes to perform in-plane transport measurements (Figure 4a). We transferred a MoS$_2$ flake on a SiO$_2$/Si substrate (Figure 4b) and made contact with the two probes to the trilayer region as illustrated in the inset in Figure 4b. The heavily doped Si layer can be used as a backgate allowing one to perform electric field effect measurements [42]. Figure 4c shows the *IV* output characteristics of the $MoS_2$ trilayer measured at different gate voltages. From these measurements one can extract the source-drain current versus back gate voltage (*I-$V_g$*) at constant bias voltage (the inset in Figure 4c shows the *I-$V_g$* for a bias voltage of 1V). The $MoS_2$ trilayer exhibits a n-type behaviour[42], as it is evidenced the current increases upon applying positive gate voltage and the fact that the channel can be switched off for negative gate voltages of the order of < -20 V. The ON-OFF current ratio obtained in our device is ~150 at bias voltage of 1V, similar to other trilayer $MoS_2$ device in literature at comparable conditions [43]. The mobility, calculated from the slope of the *I-$V_g$* trace, is between 1.3 and 3.3 $cm^2$/(V·s) (the uncertainty arises from the determination of the channel width and length), which is comparable with mobilities of 0.5 and 10 $cm^2$/(V · s) reported in the literature for trilayer $MoS_2$ devices fabricated by conventional lithography techniques [1,42–44]. Note that the uncertainty in the mobility determination, arising from the uncertainty in the channel dimensions and the contact resistance, can be reduced significantly by employing four probes and adopting a van der Pauw configuration to account for the contact resistance.

In addition to the field-effect measurements, the configuration using two carbon fiber probes ensures that the 2D material channel is fully exposed, which can be an interesting feature to fabricate proof-of-concept photodetectors or chemical sensors [45]. To demonstrate this capability of this microprobing technique to test proof-of-concept devices we show the characterization of another trilayer $MoS_2$ upon illumination [46,47].

Figure 5a shows the electrical characteristics of a device upon illumination with increasingly high power illumination. By measuring the current through the device at a fixed bias voltage while turning on and off the illumination one can characterize the response time of the photodetector, as shown in Figure 5b. In this case, we obtained for all the wavelengths rise/fall times of <100 ms (limited by our experimental setup). This value is in the range of previously reported response times, which range from μs [44,48] to s [49,50]. Additionally, Figure 5c shows the photocurrent spectra of the device measured by employing illumination sources with different wavelengths which is closely related to the absorption spectrum measured on that $MoS_2$ flake before transferring it. In the differential reflectance spectra of the $MoS_2$ device displayed in Figure 5c, there are two resonance peaks corresponding to the A and B excitons [7] that correspond to direct band-to-band transitions. There is a sudden drop of the photocurrent to zero for photons with lower energy than the A exciton (~1.85 eV) [6]. The responsivity of our sample reaches 26 A/W which is in the same range as previously reported in the literature for $MoS_2$ photodetectors (measured at similar power density ~20 mW/cm$^2$) fabricated by lithographic techniques [49].

After demonstrating the potential of the carbon fiber probes to establish electrical contact to 2D materials, we also characterize how gentle the contact with the flakes is. Therefore, we show in Figure 6 the topography of a $MoS_2$ flake deposited on a $SiO_2$/Si substrate after sweeping a carbon fiber probe along its surface (in contact). We observe in the AFM image that the carbon fiber did not scratch the surface of our flake noticeably. Note that, the same process was performed with a very thin (10 μm in diameter) metal wire probe finding that the metal probe tends to tear the flakes (see Supporting Information).

So far, all the results shown here were obtained for $MoS_2$, which is a relevant member of the 2D materials family, but we want to stress that this technique can be applied to other 2D materials. Figure 7 shows few examples of other 2D materials contacted with two carbon fiber probes (transport in-plane). We chose multilayer graphene, a representative member of 2D materials; $WS_2$, another TMDC; $TiS_3$, a transition metal trichalcogenide, and black phosphorus to cover different 2D families. We could measure *IV*s in all the samples with the carbon fiber microprobes, illustrating the adaptability of the method.

CONCLUSIONS

In this paper, we proved the suitability of carbon fibers microprobes to perform electrical measurements directly on 2D materials. Several 2D materials such as single- and few-layer $MoS_2$ multilayer graphene, $WS_2$, $TiS_3$ and black phosphorus were measured with this setup to display the versatility of the carbon fiber microprobing technique.

In conclusion, carbon fiber microprobes are a good alternative to metallic electrodes when one needs a fast, non-destructive and reversible method to perform electrical transport measurements of 2D materials.

ACKNOWLEDGEMENTS

We acknowledge funding from the European Commission under the Graphene Flagship, contract CNECTICT-604391, from EU Horizon 2020 research and innovation program under grant agreement No. 696656 (GrapheneCore1-Graphene-based disruptive technologies), from the FP7 ITN MOLESCO (project no. 606728), from the MINECO (Ramón y Cajal 2014 program RYC-2014-01406, RYC-2014-16626, MAT2014-58399-JIN and FIS2015-67367-C2-1-P), from the Comunidad de Madrid (MAD2D-CM

Program (S2013/MIT-3007)) and NANOFRONTMAG-CM program (S2013/MIT-2850). RF acknowledges support from the Netherlands Organisation for Scientific Research (NWO) through the research program Rubicon with project number 680-50-1515. YN acknowledges the grant from the China Scholarship Council (File NO. 201506120102).

FIGURES:

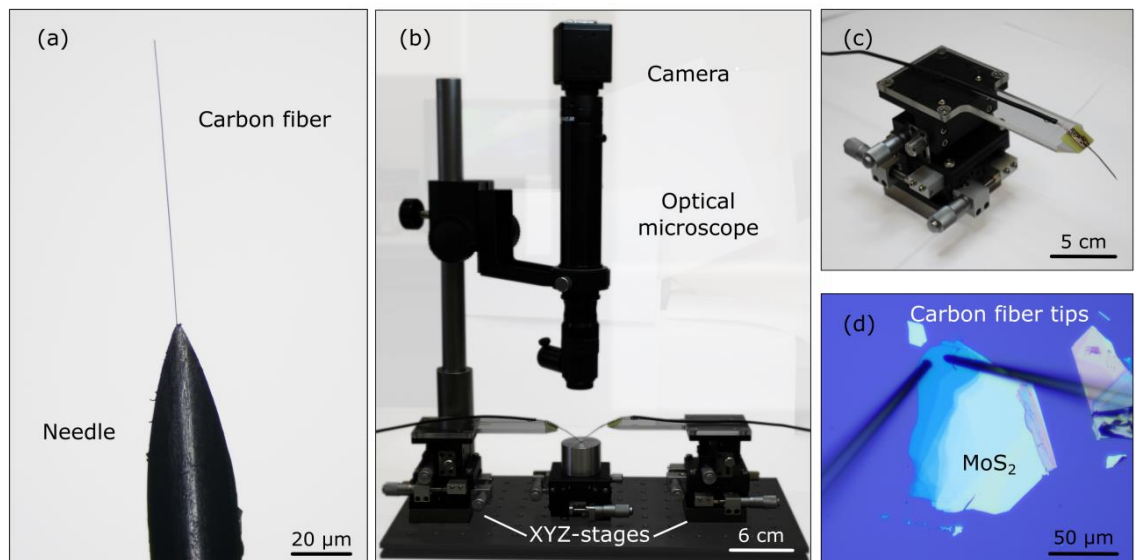

Figure 1. a) Detail of the carbon fiber tip mounted on a metallic needle that is used as microprobe. b) Picture of the carbon fiber microprobe station. c) One of the XYZ-micrometer stage used to position the probes. d) Optical microscope image of two carbon fiber tips contacting a $MoS_2$ flake transferred on $SiO_2$.

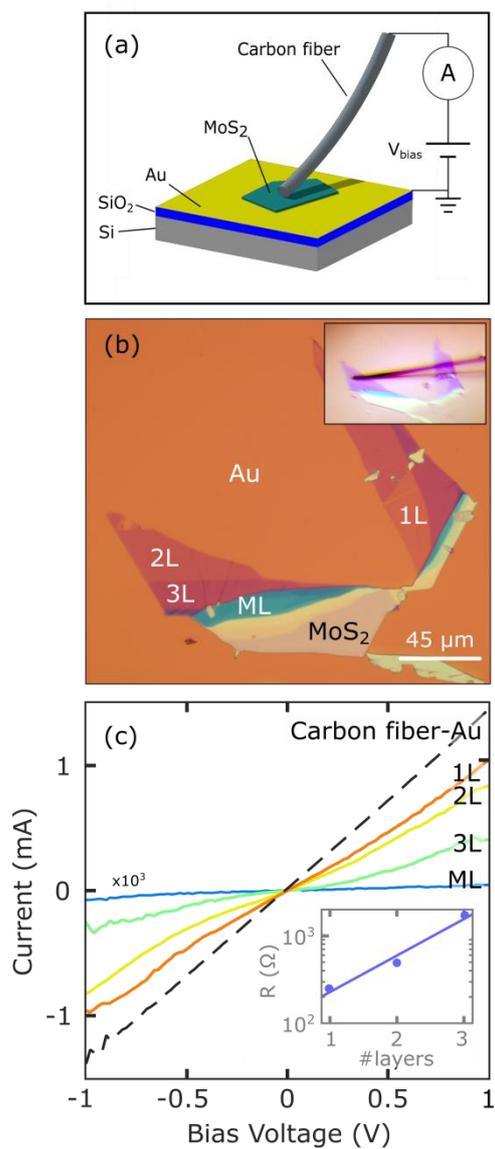

Figure 2. a) Scheme of a vertical device made out of a $MoS_2$ flake transferred on Au. b) Optical image of the sample, a $MoS_2$ flake with different number of layers transferred on Au. In the inset, the same flake contacted with one carbon fiber. c) Current-bias voltage curves measured on the different areas of the flake indicated in 2b. The inset shows the resistance dependence on the number of layers at 1 V of bias voltage. The resistance of the fiber ($R = 800 \pm 50\ \Omega$) has been taken into account.

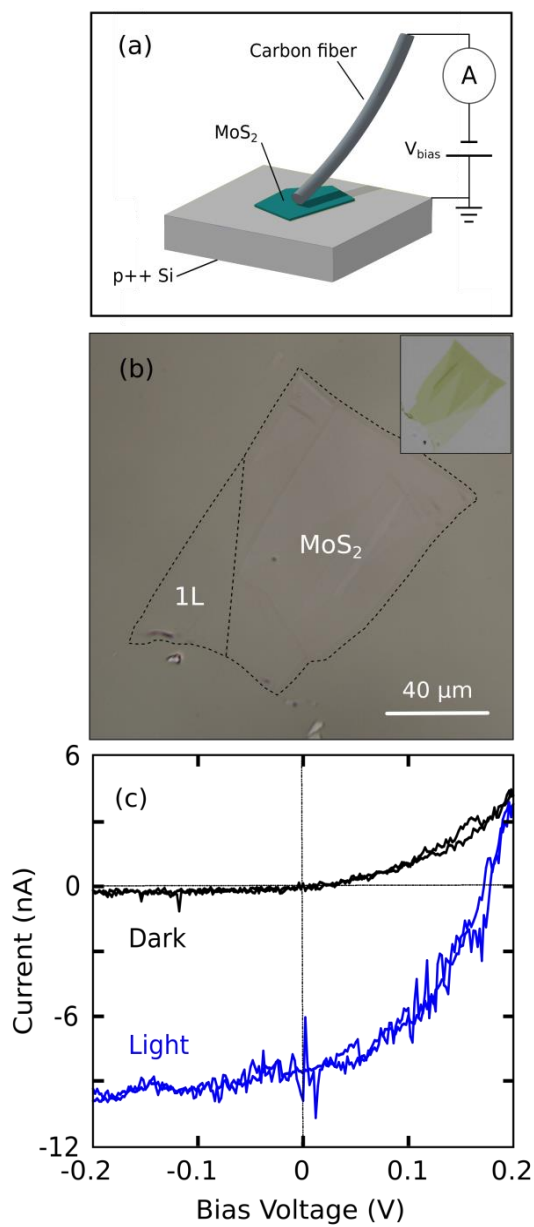

Figure 3. a) Scheme of a vertical device made out of MoS$_2$ flake, a n-type semiconductor, transferred on p-type Si. b) Optical image of the MoS$_2$ flake transferred on Si. In the inset, the same MoS$_2$ flake on PDMS before to be transferred.
c) Experimental output (*IV*) characteristics of the p-n junction, rectification in the dark (black) and $V_{oc}$ and $I_{sc}$ under illumination (white light).

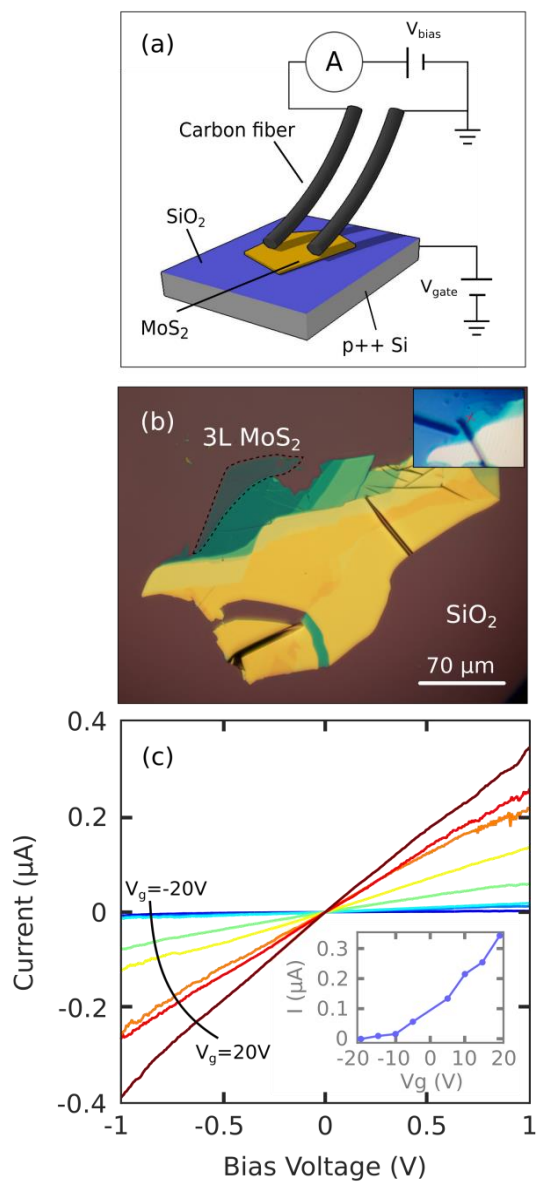

Figure 4. a) Scheme of a $MoS_2$ flake on a $SiO_2$/Si substrate and contacted with two carbon fibers. b) Optical image of the $MoS_2$ sample. The inset shows the same flake contacted with two carbon fibers. c) Current-bias voltage curves measured at different gate voltage. The inset shows the current dependence of the gate voltage for a bias voltage of 1V, extracted from the *IV*s.

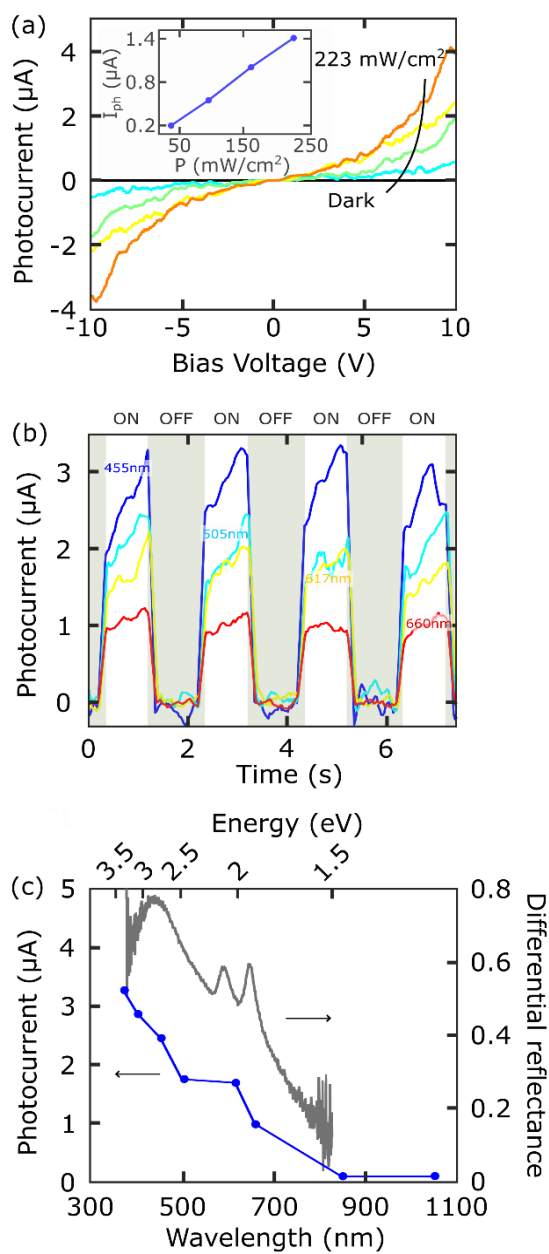

Figure 5. a) Photocurrent vs. bias voltage curves measured under illumination (455 nm) at different powers. The inset shows the photocurrent dependence on the power for bias voltage of 10 V. b) Response time for 660 nm, 617 nm, 505 nm and 455 nm of illumination wavelength with power density of 160 mW/cm$^2$ and bias voltage of 10 V. c) Photocurrent spectra measured for different illumination wavelengths with a power density of 140 mW/cm$^2$ and bias voltage of 10 V. The photocurrent spectra is compared with the differential reflectance spectrum.

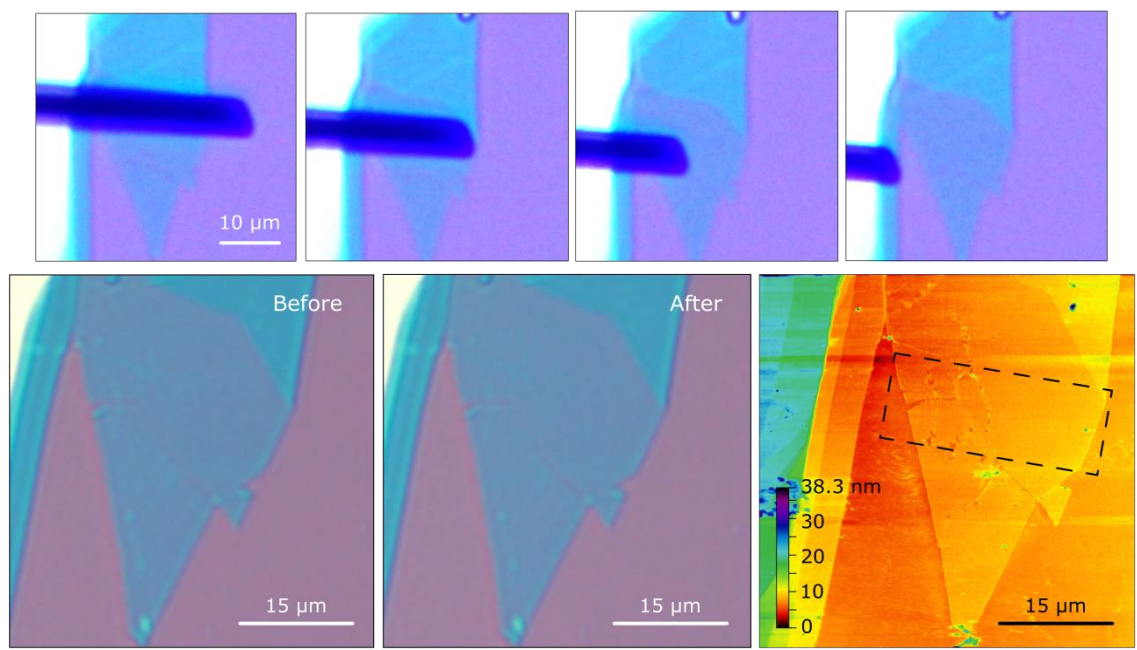

Figure 6. In the first row, four optical images of one carbon fiber sweeping in contact along a monolayer $MoS_2$ flake. In the second row, pictures taken before and after sweeping with the carbon fiber probe and AFM image after brooming the carbon fiber.

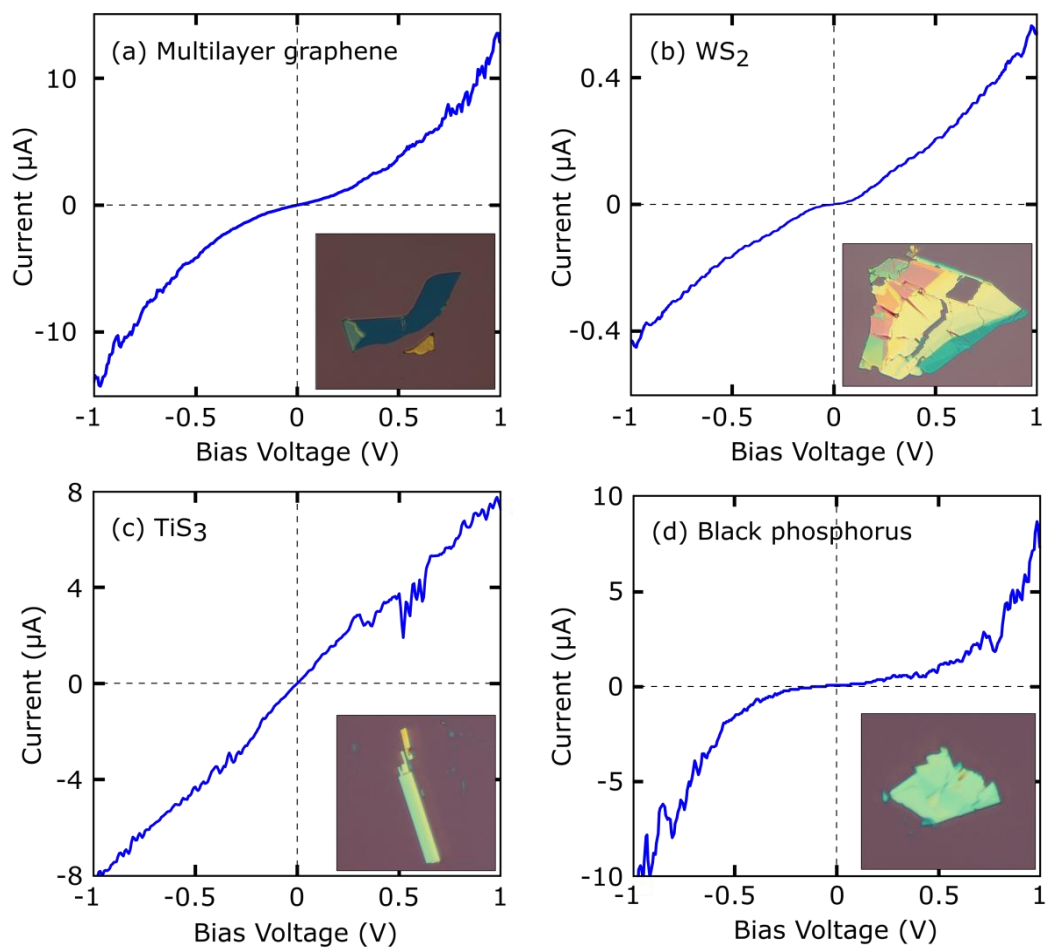

Figure 7. Current versus bias voltage curves for a a) graphite flake, a b) WS$_2$ flake, c) TiS$_3$ flake and d) black phosphorus flake.

# Supporting Information: Lithography-free electrical transport measurements on 2D materials by direct microprobing

*Patricia Gant[1], Yue Niu[1,2], Simon A. Svatek[1,3], Nicolás Agraït[1,3], Carmen Munuera[4], Mar García-Hernández[4], Riccardo Frisenda[1], David Perez de Lara[1], Andres Castellanos-Gomez[4,]*\**

[1] *Instituto Madrileño de Estudios Avanzados en Nanociencia (IMDEA Nanociencia), Campus de Cantoblanco, E-28049 Madrid, Spain.*

[2] *National Key Laboratory of Science and Technology on Advanced Composites in Special Environments, Harbin Institute of Technology, Harbin, China*

[3] *Departamento de Física de la Materia Condensada and IFIMAC. Universidad Autónoma de Madrid, Madrid, E-28049, Spain.*

[4] *Instituto de Ciencia de Materiales de Madrid (ICMM-CSIC), Madrid, E-28049, Spain.*

andres.castellanos@csic.es

**Comparison between a human hair and a carbon fiber**

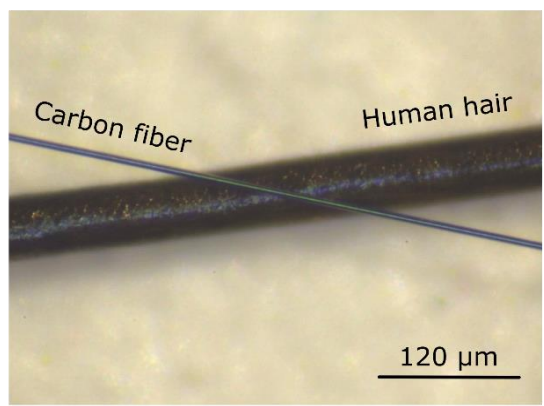

Figure S1. Optical image of a human hair and a PAN carbon fiber (7 μm in diameter).

**Assembling carbon fiber microprobes**

Figure S2 shows pictures of the different steps required to assemble the carbon fiber microprobes. A new probe can be fabricated in 20-30 minutes (15-20 minutes are required to let the silver lodaded adhesive dry).

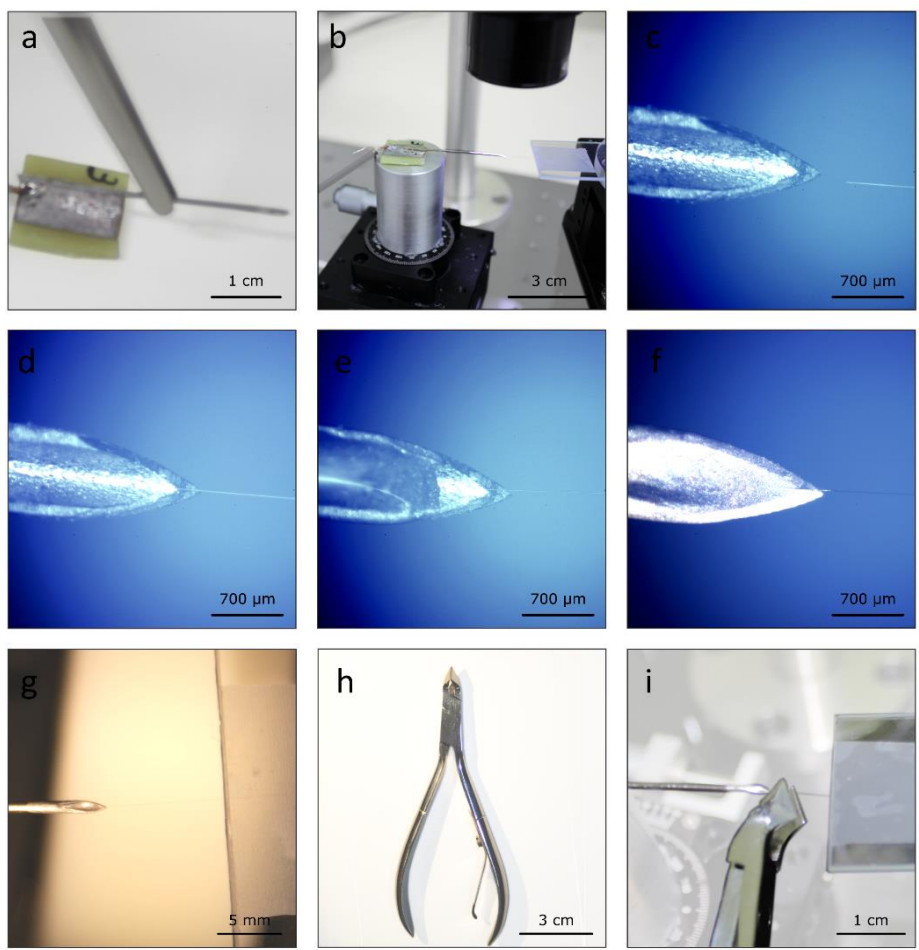

Figure S2. Pictures of the different steps employed in order to assemble the carbon fiber microprobes. (a) A surgical steel needle is soldered to a printed circuit board (PCB). (b) The PCB with the steel needle is mounted in a XY micro-manipulator stage and a long carbon fiber is brought close to the needle with a XYZ micro-manipulator stage. The process is carried out under optical inspection with a long working distance zoom lens (like the one shown in Figure 1b). (c) Zoomed in image of the steel needle and the carbon fiber end, as seen through the zoom lens. (d) The fiber is placed in contact to the metal needle (with more than 1 mm of overlap between the metal needle and the carbon fiber). (e) A small droplet of silver loaded glue is placed in the needle. (d) Image of the carbon fiber glued to the metal needle after letting the silver glue drying. (g) The carbon fiber is attached to both the metal needle and to the XYZ stage and it has to be cut to the desire length using a nail cuticle nipper (h) and (i).

## Characterization of the carbon fiber microprobes

### *Mechanical properties*

The spring constant ($k$) of the carbon fiber microprobes can be calculated by considering them cylindrical cantilevers [S1]:

$$k = \frac{r^4 \cdot E}{2 \cdot L^3 \cdot (1 - v^2)}$$

where $r$ is the fiber radius, $E$ is the Young's modulus, $L$ is the length and $v$ is the Poisson's ratio. Using the values reported in Ref. [S1] for the Young's modulus ($E = 246 \pm 8$ GPa) and the Poisson's ratio ($v = 0.27$) the spring constant for a typical carbon fiber microprobe ($L = 1$ mm) is $k \sim 0.02$ N/m. Therefore, the carbon fiber microprobes are very floppy, allowing very gentle mechanical contacts to the flakes without damaging them.

One can also estimate the mechanical resonance frequency of the microprobes as a very low mechanical resonance (in the order of 1-100 Hz) can yield to mechanical instability of the probes due to excitation with acustic noise and lab vibrations.

The resonance frequency ($f$) of a cylindrical cantilever is given by [S1]:

$$f = \frac{1}{2\pi}\sqrt{\frac{k}{m_{\text{eff}}}}$$

where $m_{\text{eff}}$ is the effective mass of the carbon fiber ($m_{\text{eff}} = 0.243 \cdot \rho \cdot L \cdot \pi r^2$). This gives a resonance frequency of ~5.6 kHz, which is well-above the human voice spectral range (up to 3.4 kHz) and thus it guarantees higher mechanical stability.

*Electrical properties*

*Determination of the carbon fiber electrical resistivity and contact resistance*

Following the procedure described in Ref. [S1] we place a single carbon fiber onto a glass slide and put drops of silver loaded adhesive contacts along the fiber. The contact electrical resistivity of the fiber was obtained by measuring the resistance of the carbon fiber between the different contacts. The relationship between the resistance and the distance between the silver paint contacts is given by:

$$R = (\rho_{el} A^{-1}) \cdot L + 2 \cdot R_C \qquad (1)$$

where $R$ is the total resistance, $\rho_{el}$ the resistivity of the carbon fiber, $A$ the cross section of the fiber, $L$ the length of the fiber between two silver paint contacts and $R_C$ the contact resistance between the fiber and the silver paint. Figure S3 shows the relationship between the measured electrical resistance and the distance between the silver paint contacts, following a straight line. The slope of this straight line allows us to calculate the electrical resistivity of the carbon fiber, which is $1.92 \cdot 10^{-5}$ $\Omega$m.

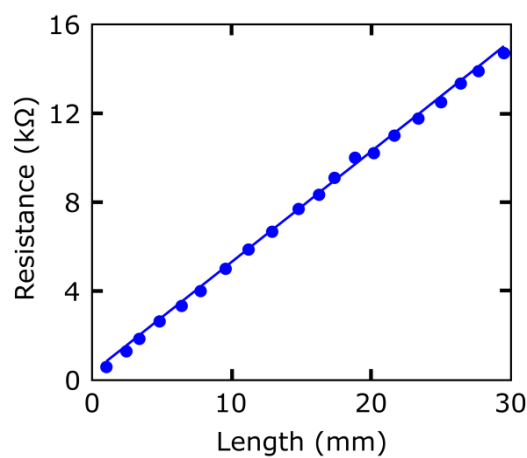

Figure S3. The resistance measured when a carbon fiber probe contacts with Au substrate for different lengths of the carbon fiber.

The interception with the vertical axis of the straight line allows us to extract the contact resistance between the carbon fiber and the silver loaded adhesive, also used to attach the carbon fibers to the metallic needles, $R_C = 80\ \Omega$.

*Determination of the contact resistance between the carbon fiber and a 2D semiconductor*

In this case, we used a flake of TiS$_3$ to determine the contact resistance with a 2D semiconductor. We chose this material due to its ribbon shape geometry, which ensures a homogeneous channel width and allows us to easily measure the channel length. The contact resistance was obtained again by using Expression (1).

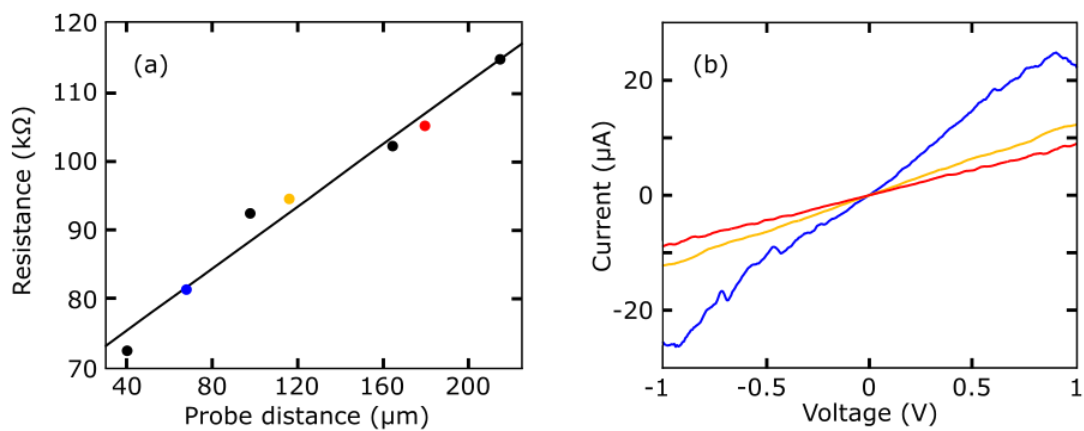

Figure S4. a) Resistance measured when we used two carbon fiber probes to contact a TiS$_3$ flake, changing the distance between the carbon fiber tips. b) Current vs. voltage curves used to extract the resistance at different probe distance.

The contact resistance value obtained from the linear fit is 33.1 kΩ (or 8.3 Ω·cm). This contact resistance is in the same range that the contact resistance between the trilayer $MoS_2$ and Au electrodes reported in Reference [S2].

**Detecting the contact by monitoring the probe-sample current during the probe approaching**

Figure S5 shows the current monitored while the probe is lowered to make an electrical contact to the sample. A sudden increase in the current indicated that the contact is established.

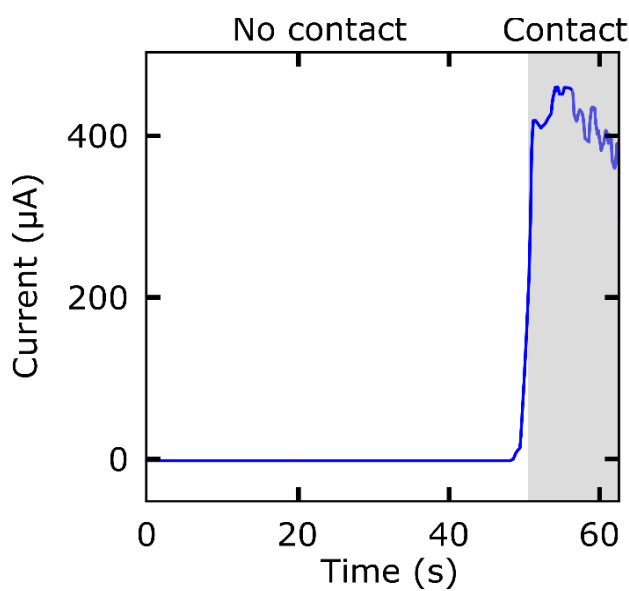

Figure S5. Monitoring the current while the probe is approached to the surface allows to know when the carbon fiber is in contact with the gold substrate. The fluctuation of the current in the contact region is due to the manual actuation of the micromanipulator.

**Reproducibility and stability of the carbon fiber microprobe measurements**

Unlike conventional micro/nano lithographic contacts, carbon fiber microprobe contacts are movable contacts and thus one has to characterize their reproducibility and stability in time.

Figure S6a shows 4 sets of *IV*s measured on a multilayer $MoS_2$ flake on gold with a carbon fiber (vertical configuration, similarly to Figure 2 in the main text). In each *IV* both trace and retrace have been measured, showing an excellent reproducibility. Figure 6b shows the resistance extracted from the *IV*s measured for flakes with different number of layers (the filled circles correspond to the traces and the empty ones to the retraces). This measurement illustrates the dispersion from measurement to measurement which we attribute to be due to mechanical vibrations of the different part of the experimental setup. Note that all the measurements were done on top of a normal laboratory desk. Therefore, the stability of the measurements could be further improved by using a vibration isolation table and a acustic isolated enclosure for the probe station.

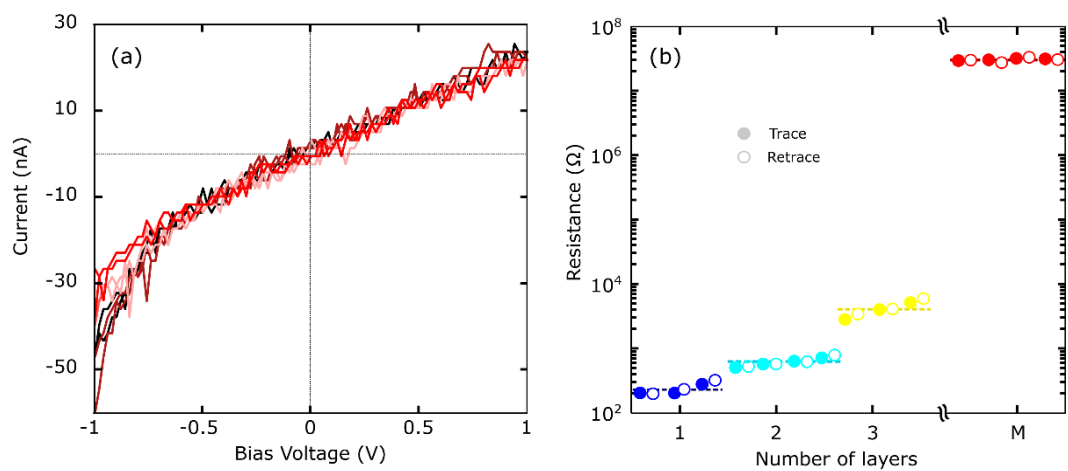

Figure S6. (a) Current *vs*. voltage characteristics measured on a multilayer MoS$_2$ flake in vertical configuration, for each IV set both trace and retrace are shown. (b) Electrical resistance extracted from IVs measured for MoS2 flakes with different thicknesses

Longterm timetrace measurements can also provide an insight about the estability of the carbon fiber microprobes. Figure S7 shows timetraces of the photodetector device shown in the Figure 5 of the main text. The Figure illustrates how the measurement can be stable at time scales up to 10-20 seconds but at certain point the whole trace drifts. We attribute these drifts due to air flow that leads to a differential thermal drift of the parts of our setup. This could be improved substantially by enclosing the probe station in a airtight box.

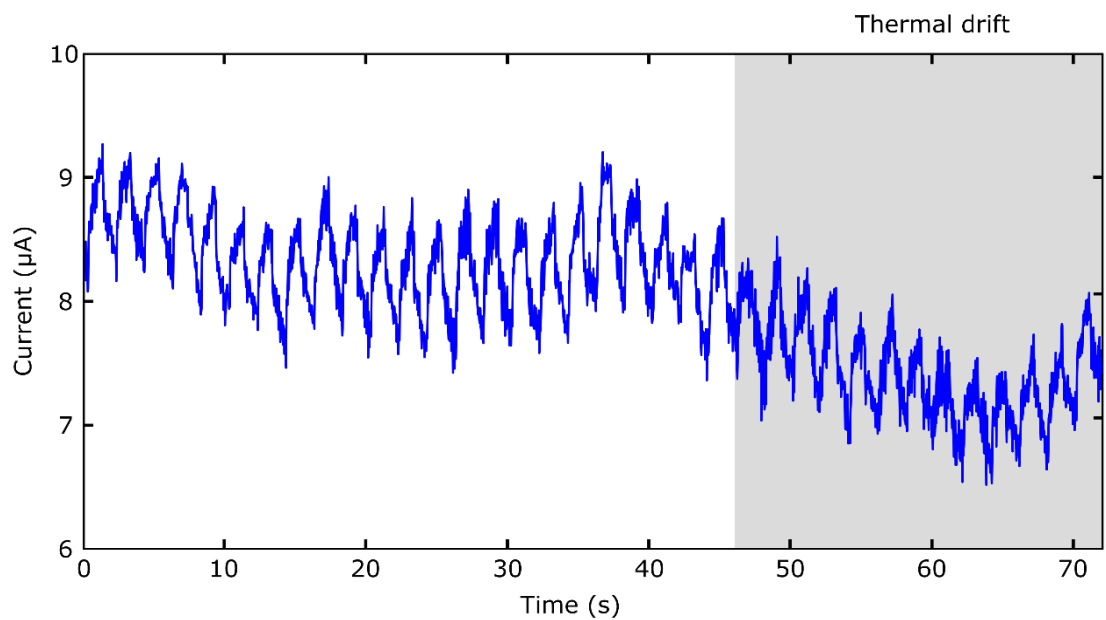

Figure S7. Time traces of the photodetector MoS$_2$ device shown in Figure 5. The drain source current is being measured at a fixed bias voltage of 10 V while the illumination is being switched on and off. Power density of 160 mW/cm$^2$ and wavelength 505 nm.

**Differential reflectance spectra of MoS$_2$ flakes**

In order to determine the thickness of the different areas of the studied MoS$_2$ flake, we measured the differential reflectance spectrum, previously calibrated, with the same method described in Reference [S3]. We show in Figure S8 the different spectra obtained for 1L, 2L and 3L in comparison with the calibrated spectra.

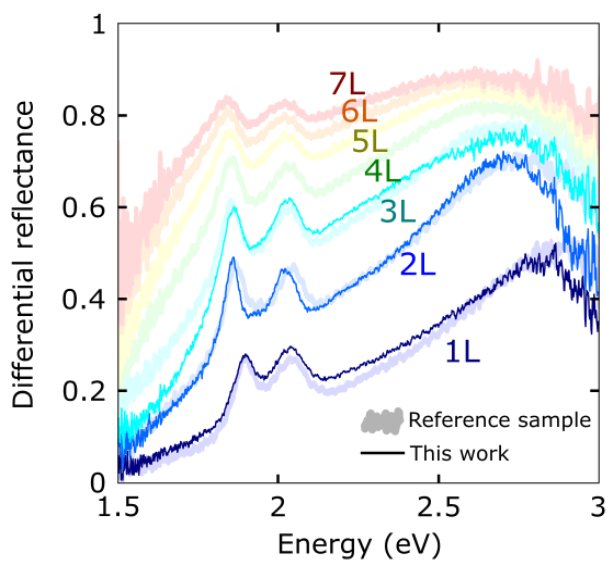

Figure S8. Shading lines are the differential reflectance spectra of the reference sample and the three overlapping spectra show the differential reflectance of 1L, 2L and 3L from the sample used in this work.

**Carbon fibers vs. thin metallic wire**

The reason of choosing the carbon fibers and not using metallic wires, despite this higher contact resistance and resistivity, is the gentle contact of carbon fibers. We also tried to use thin metallic wires to contact the flakes, but the wires tend to scratch the surface of the flakes, as Figure S9 shows. The metallic wire chosen was a Pt Wollaston wire with 10 μm of diameter. This difference might be due to the presence of sharp asperities in the metal wire surface as compared with the carbon fiber surface that tends to be smoother (see Figure S10) or to the fact that carbon fibers are known to slide against surfaces with a rather small friction coefficient (as compared with metals) [S4].

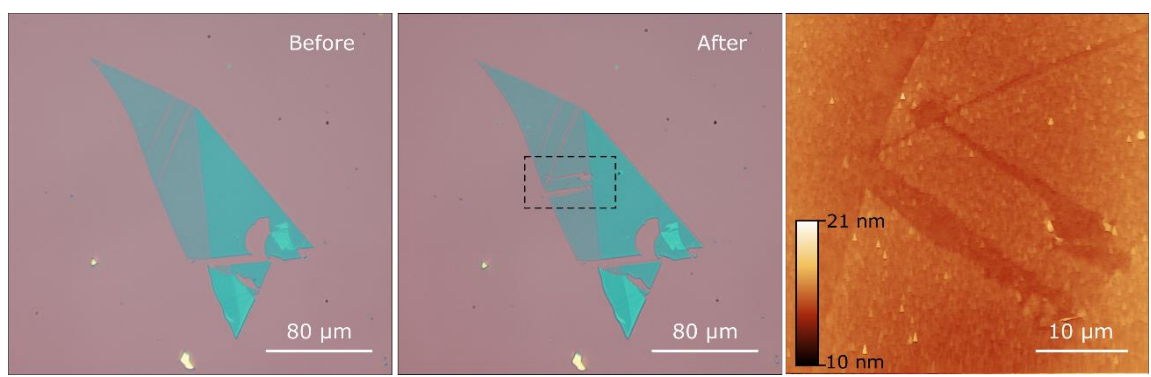

Figure S9. Optical images of a $MoS_2$ flake transferred on $SiO_2$ before and after sweeping with the Wollaston wire. The AFM image shows the scratches made by the wire.

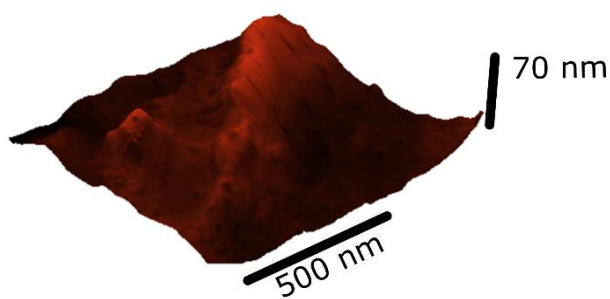
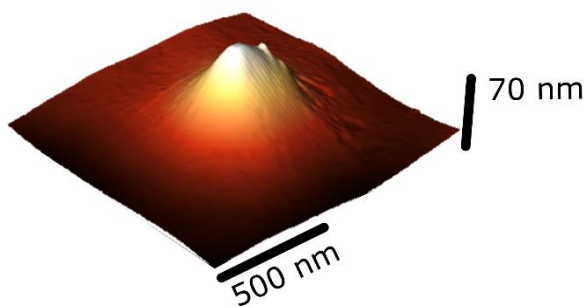

Figure S10. AFM topography images of the surfaces of a carbon fiber and a Wollaston wire.

**More 'in-plane measurements'**

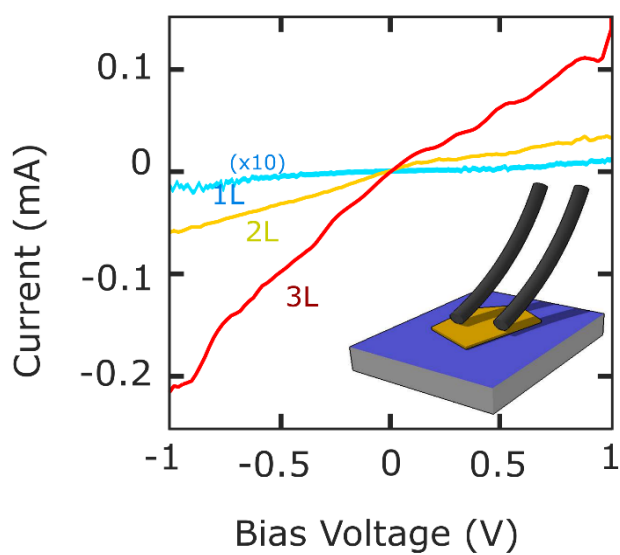

Figure S11. Current-bias voltage curves measured in 'in-plane' configuration, using two carbon fiber probes on $MoS_2$ flakes with different number of layers as indicated by the labels monolayer (1L), bilayer (2L) and trilayer (3L). Note that the current of the monolayer has been multiplied by 10 to facilitate the comparison with the bilayer and trilayer flakes. All the measurements have been carried out at $V_{gate} = 0V$.

**Supporting Information references:**